\documentclass[11pt]{article} 

\usepackage[utf8]{inputenc} 

\usepackage{amssymb}

\usepackage{geometry} 
\usepackage{graphicx} 

\usepackage{booktabs} 
\usepackage{array} 
\usepackage{paralist} 
\usepackage{verbatim} 
\usepackage{subfig} 

\usepackage{fancyhdr} 
\usepackage{sectsty}
\allsectionsfont{\sffamily\mdseries\upshape} 

\usepackage[nottoc,notlof,notlot]{tocbibind} 
\usepackage[titles,subfigure]{tocloft} 

\title{One class of linear Fredholm integral equations with  functionals and parameters}
\author{L.R. Dreglea Sidorov, N. Sidorov and  D. Sidorov}

\begin{document}
\maketitle

\section{Introduction}

This paper deals with some issues in the theory of linear integral equations with linear functionals.
Modern views on the fundamental laws of nature are often stated in terms of integral equations~\cite{Azbelev, Hrom, Nah,Sid, Sid1}.
 The analysis of such operators includes questions of finding eigenvalues and adjoint  functions \cite{Vain}, studying the convergence of their asymptotics, existence and convergence theorems of approximate methods \cite{Sid, Sid1}. At the end of 20th century, A. P. Khromov found a new class of integral operators with discontinuous kernels and began a systematic study of them \cite{Hrom}.
 Under very general assumptions, he derived the conditions under which eigenfunction expansions of these operators behave like trigonometric Fourier series. However, these conditions as well as the construction of the classical discontinuous Fredholm resolvent in the form of the ratio of two integer analytic expansions over a parameter are difficult to verify.
 In the works \cite{Sid13, Sid14, Sid1} a class of equations with discontinuous kernels was distinguished and studied.

In \cite{Math22} the branching solutions of the Cauchy problem for nonlinear loaded differential equations with bifurcation parameters were studied. 
 The purpose of this study is to prove the properties of the resolvent integral operator as applied to the second kind Fredholm integral equations with local and integral loads, and to formulate and prove constructive theorems of existence and convergence to the desired solution of successive approximations.

 Let us consider the equation
 \begin{equation}
 x-{\mathcal L}x - \lambda {\mathcal K} x = f,
 \label{eq1} 
 \end{equation}
where linear operators ${\mathcal L}$ and ${\mathcal K}$ are given as follows
$$ {\mathcal L}x := \sum_{k=1}^n a_k(t) \langle \gamma_k, x \rangle, $$
$$ {\mathcal K}x := \int_a^b K(t,s) x(s)\, ds, $$
$\lambda$ is parameter. All the functions in  (\ref{eq1}) are continuous. Kernel $K(t,s)$ can be symmetric  and it is also continuous both in $t$ and $s$.
The desired solution $x(t)$ is constructed in ${\mathcal C}_{[a,b]}.$

Linear functionals $\langle \gamma_k, x \rangle$ in applications corresponds to 
the {\it loads} imposed on the desired solution. The loads can be local $ (\langle \gamma_k, x \rangle = x(t_k), \, t_k \in [a,b])  $ or integral such as 
$\langle \gamma_k, x \rangle = \int_a^b \gamma_k(t) x(t)\, dt,$
where $\gamma_k(t)$ are piecewise continuous functions  for $t\in [a,b]$
or  $\langle \gamma_k, x \rangle = \int_a^b  x(t)\, d \gamma_k(t),$
$\gamma_k(t)$ is given function of limited variation.

The objective is to construct solution $x(t,\lambda)$ for $\lambda \in {\mathbb R}^1$
of equation  (\ref{eq1}).  For operator ${\mathcal L}x$ below the following brief notation 
$$ {\mathcal L}x  :=  \sum_{k=1}^n a_k(t) \langle \gamma_k, x \rangle \equiv (\vec{a}(t), \langle \vec{\gamma}, x \rangle ) $$ is used, where conventional notation  $(\cdot, \cdot)$ for scalar product is used. Here $\vec{a}(t) = (a_1(t), \cdots, a_n(t))^T,$ $a_i(y) \in {\mathcal C}_{[a,b]},$  $\langle \vec{\gamma}, x \rangle  = ( \langle {\gamma_1}, x \rangle, \dots, \langle {\gamma_n}, x \rangle  )^T.$

Loaded differential equations have been intensivly studied during the last decades.
The term ``loaded equation'' was first  used in the works of A.M. Nakhushev, here readers may refer to his monograph \cite{Nah}.
Loaded equations appears in many applications, see e.g. \cite{Chad, Bal}
But theory and numerical methods for the loaded integral equations remained less developed.
In paper \cite{SidSid} the problem statement for the integral equation with single load is given.
Then, in \cite{Sid23, SidDr}  theory of the Hammerstein integral equations with loads and bifurcation parameters was proposed. In \cite{Lampe} the Fredholm resolvent was employed for computing $H_2$-norm for linear periodic systems. 

 The similar statement is addressed in the present paper and analytical method is describled 
 which makes it possible to consider integral equations with arbitrary finite number 
 of local and integral loads. An example of functionals that generate local and integral loads in the space  ${\mathcal C}_{[a,b]}$ is the functional
$$  \langle \gamma, x \rangle :=  \sum_{i=1}^m \alpha_i x(t_i) + \sum_{i=1}^n \int_{a_i}^{b_i} m_i(s)
x(s)\, ds, $$ 
where $\alpha_i \in {\mathbb R}^1,$ $[a_i, b_i] \subset [a,b],$ $m_i (s) \in {\mathcal C}_{[a_i, b_i]},$  $t_i \in [a,b].$ 

\section{System of equations to determine the load}

Let us introduce the following condition

\noindent {\bf I.} $\langle \gamma_k, K(t,s) \rangle = 0, \, k=1,\dots, n, $  $s \in [a,b]$ and  vectors
$ \vec{x}_\gamma = (\langle \gamma_1, x \rangle, \dots, \langle \gamma_n, x \rangle)^T, $
$ \vec{f}_\gamma = (\langle \gamma_1, f \rangle, \dots, \langle \gamma_n, f \rangle)^T. 
$


\noindent {\it Lemma 1}\\
Let condition {\bf I} be fuilfilled. Then load vector $\vec{x}_j$ necesserily satisfies system 
 (\ref{eq2})  
\begin{equation}
(E-A_0)\vec{c} = \vec{f}_{\gamma},
\label{eq2}
\end{equation}
  where $A_0 =  [\langle \gamma_i, a_k \rangle ]_{i,k=1}^n,$ $E$ is $(n \times n)$ identity matrix.\\

\noindent Proof. \\
Let us apply the functionals $\langle \gamma_i, \cdot \rangle,$  $i=1, \dots, n$
to both parts of equation  (\ref{eq1}). Using  {\bf I}, the following system can be derived
\begin{equation}
\langle \gamma_i, x \rangle - \sum_{k=1}^n \langle \gamma_i, a_k \rangle  \langle \gamma_k, x \rangle = \langle \gamma_i, f \rangle, \, i=1,\dots, n.   
\label{eq3}
\end{equation}
System of linear algebraic equations  (\ref{eq3}) is in fact, system  (\ref{eq2})
presented in coordinate system. Lemma is proved.

From this Lemma it follows:

\noindent {\it Corollary 1}\\
Let condition {\bf I} be fuilfilled and system  (\ref{eq2})  has no solution. Then equation 
 (\ref{eq1}) has no solution in class of continuous functions. 

Let condition {\bf I} be fuilfilled and vector $\vec{c}^* \in {\mathbb R}^n$ satisfies system 
 (\ref{eq2}). Then solution $x(t, \lambda)$ of equation  (\ref{eq1})
depends on vector $\vec{c}^*$ satisfies the following Fredholm integral equation of the 2nd kind
$$
x(t,\lambda) - \lambda \int_a^b K(t,s) x(s,\lambda)\, ds = f(t) + (\vec{a}(t), \vec{c}^*).
$$

\noindent {\it Lemma 2}\\
\label{lem2}
Solution of equation  (\ref{eq1}) for arbitrary $\lambda,$ expect  the characteristic numbers
$\lambda_i$ of kernel $K(t,s),$ is defined by following formula 
\begin{equation}
x(t,\lambda) = (\vec{a}(t), \vec{x}_{\gamma}(\lambda)) + \int_a^b \Gamma(t,s,\lambda)
(\vec{a}(s), \vec{x}_{\gamma}(\lambda))\, ds + \int_a^b  \Gamma(t,s,\lambda) f(s)\, ds + f(t).
\label{eq4}
\end{equation}
Here $\Gamma(t,s,\lambda) = \frac{D(t,s,\lambda)}{D(\lambda)},$
$D(t,s,\lambda)$ and $D(\lambda)$ are the entire analytic functions of parameter $\lambda,$
$D(\lambda_i) =0.$ Load vector $\vec{x}_{\gamma}(\lambda)$ is necesserily must 
satisfy the following system of  $n$ linear algebraic equations 
\begin{equation}
(E-A_0-A(\lambda)) \vec{x}_{\gamma}(\lambda) = \vec{b}(\lambda)
\label{eq5}
\end{equation}
with matrix 
\begin{equation}
A(\lambda) = \biggl  \langle  \gamma_i, \int_a^b \Gamma(t,s,\lambda) a_k(s)\, ds \biggr \rangle_{i,k=1}^n
\label{eq6}
\end{equation}
and vector $$ \vec{b}(\lambda) =   \biggl  \langle  \gamma_i, f(t)+\int_a^b \Gamma(t,s,\lambda) f(s)\, ds \biggr \rangle_{i=1}^n.  $$
The set of characteristic numbers $\{ \lambda_i \}$ is finite  and countable set .\\

\noindent Proof.\\ 
It is known (see sec. 9 (3) in book \cite{Kolm}) that an inverse operator $(I-\lambda K)^{-1}$
is defined by Fredholm formula \cite{Fred}:
$$(I-\lambda K)^{-1} = I+ \lambda \int_a^b \frac{D(t,s,\lambda)}{D(\lambda)} [\cdot]\, ds.$$
Functions $D(t,s,\lambda)$ and $D(\lambda)$ are entire analytical funcations 
with respect to $\lambda,$ defined for $\lambda \in {\mathbb R}^1.$
Moreover, characteristic numbers of kernel $K(t,s)$ of operator ${\mathcal K}$
are nuls of denumerator $D(\lambda).$
 Thus, inverse operator $(I-\lambda K)^{-1}$ can be called as discontinuous operator.
 Indeed, function $\Gamma(t,s,\lambda)$ in solution  (\ref{eq4}) has 
 the 2nd kind discontinuities in points $\{\lambda_i\}.$
 By solving system  (\ref{eq5}) and substituting its solution into  (\ref{eq4}), we find the solution of the original problem  (\ref{eq1}). The lamma is proved.\\

\noindent {\it Remark 1}\\ 
In system  (\ref{eq5}) in general case matrix $A(\lambda)$ and vector $\vec{b}(\lambda)$
will have 2nd kind discontinuities in points $\lambda.$

Let us  distinguish the class of  kernels $K(t,s)$ when matrix $A_0$ and vector $\vec{b}(\lambda)$ can be specified.
 Let the kernel $K(t,s)$ generate the nilpotency of the operator ${\mathcal K}.$
 
 Let  $|\lambda|<\frac{1}{||{\mathcal K}||}.$ In that case solution of equation 
 $x-\lambda {\mathcal K}x = f$ for arbitrary source function $f$ is defined uniqly as
 follows 
 $$x = f+\lambda {\mathcal K} f + \lambda^2 {\mathcal K^2} f + \dots + \lambda^p 
 {\mathcal K}^p f.   $$
 Here $${\mathcal K}^n f = \int_a^b K_n(t,s) f(s)\, ds, $$ where 
 $$K_n(t,s) = \int_a^b K(t,z) K_{n-1}(z,s)\, dz. $$ Here $K_1(t,s) := K(t,s),$
 $K_{p+1}(t,s) = 0$ due  to the nilpotency of the operator ${\mathcal K}$
 for some $p\geq 1.$
 Therefore, formula  (\ref{eq4}) can be presented in the following constructive form
 \begin{equation}
 x(t,\lambda,\vec{x}_{\gamma}) = f(t) +(\vec{a}(t), \vec{x}_{\gamma}) + \int_a^b \bigl (
 \lambda K(t,s) +\lambda^2 K_2(t,s) + \cdots + \lambda^p K_p(t,s)
 \bigr ) (f(s) + (\vec{a}(s), \vec{x}_{\gamma} ))\, ds.
  \label{eq7}
 \end{equation}
 Correspondingly, we derive the refined system of linear algebraic equations 
  (\ref{eq5}) with respect to the load vector because  
 \begin{equation}
 A(\lambda) = \bigl \langle  \gamma_i, \int_a^b(\lambda K(t,s) +\lambda^2 K_2(t,s) + \dots + 
 \lambda^p K_p(t,s) )  a_k(s) \, ds \bigr \rangle_{i,k=1}^n 
 \label{eq8}
 \end{equation}  

 \begin{equation}
\vec{b}(\lambda) = \bigl \langle  \gamma_i, \, f(t)+ \int_a^b(\lambda K(t,s) +\lambda^2 K_2(t,s) + \dots + 
 \lambda^p K_p(t,s) )  f(s) \, ds \bigr \rangle_{i=1}^n. 
 \label{eq9}
 \end{equation}  
 Thus, $A(\lambda)$ and $\vec{b}(\lambda)$ are continuous on $\lambda.$
It is to noted that if $\langle \gamma_i , K(t,s) \rangle = 0,\, i=1,\dots, n $ then $A(\lambda) = 0,$ and system  (\ref{eq5}) is  degenerates into system   (\ref{eq2}) introduced in Lemma 1.
Therefore, in this case vector $\vec{x}_{\gamma}$ from solution  (\ref{eq7}) for given problem  (\ref{eq1}) can be determined.  
 The following theorem follows from this. \\
 
 \noindent {\it Theorem 1.}\\
 Let operator ${\mathcal K}$ be nulpotent and $\langle \gamma_i, K(t,s) \rangle = 0,$
 $i=1, \dots, n, \, \forall s\in [a,b].$ Then exsts solution of equation  (\ref{eq1}) as 
 functional polynomial  (\ref{eq7}) of $p$-th order on parameter $\lambda$.  
 Coefficients of polynomial  (\ref{eq7}) depends on selection of the load vector 
 $\vec{x}_{\gamma}$ in ${\mathbb R}^n$. \\
  
 If operator ${\mathcal K}$ is not nulpotent and  an identity $\langle  \gamma_i, K(t,s)  \rangle = 0 $ is not satisfied, then solution $x(t,\lambda)$ of equation  (\ref{eq1}) can be found in the class of continuous on $t$ functions. This solution can be represented  in the punctured neighborhood $0<|\lambda|< \rho$ in the form of Laurent series with pole in point $\lambda=0.$
 
 \section{Successive approximations}

 Let $\det (E-A_0) \neq 0.$ Then exists  neighborhood of  $\lambda$  $|\lambda|<\rho$
such as system  (\ref{eq5})  enjoys solution $\vec{x}_{\rho}(\lambda) \rightarrow (E-A_0)^{-1}
\vec{f}_{\gamma}$ as $\lambda \rightarrow 0.$
Positive $\rho$ exists since $||(E-A_0)^{-1}A(\lambda)|| \rightarrow 0$ as $\lambda \rightarrow 0.$

Let us call the case of $\det (E-A_0) \neq 0$ as regular.\\

\noindent {Theorem 2.}\\
In the regular case $\det (E-A_0) \neq 0$ there exists neighborhood $|\lambda| < \rho$
in which equation  (\ref{eq1}) has the unique solution continous on $t$ and holomorphic on
$\lambda.$  
 
\noindent{\it Corollary 2}\\ 
Let $\det (E-A_0) \neq 0,$  $|| (I-L)^{-1}{\mathcal K} ||\leq l. $
Fix the scalar $q<1.$
Then for $|\lambda| \leq \frac{q}{l}$ equation  (\ref{eq1}) has unique solution. Moreover, solution 
 is golomorphic on $\lambda.$ The sequence $\{ x_n \},$
 where $x_n  = \lambda(I-L)^{-1}{\mathcal K} x_{n-1} + (I-L)^{-1 f}, \, x_0=0,$
 uniformly converges to the desired solution $x(t,\lambda)$ of equation  (\ref{eq1})
at the rate of a geometric progression with the denominator $q<1.$ 

Let us focus now  the irregular case of $\det (E-A_0)=0.$
Let $A_0=E.$ Then $\det (E-A_0) = 0$ and we have irregular case.
Let $\frac{d^i}{d\lambda^i} A(\lambda)\bigr|_{\lambda=0}$ for $i=0,1,\dots , p-1 $
are zero matrices and $\frac{d^p}{d\lambda^p} A(\lambda)\bigr|_{\lambda=0} \neq 0.$ 
Then load vector $\vec{x}_{\gamma}$ satisfies the following system
$$ \bigl(-E-A_p^{-1} \sum_{m=p+1}^{\infty} \lambda^{m-p} A_m \bigr)  
\vec{x}_{\gamma} = \lambda^{-p}A^{-1}\vec{b}(\lambda),$$
where $$A_{p}=\frac{1}{p!} \bigl(\frac{d^p}{d \lambda^{p}} A(\lambda)\bigr)\bigr|_{\lambda=0}. $$
Let's select neighborhood $|\lambda|<\rho$ such as $$ || A_p^{-1} \sum_{m=p+1}^{\infty}
\lambda^{m-p} A_m  || \leq q<1. $$

Then

$$ \lambda^p \vec{x}_{\gamma} = -  \sum_{n=0}^{\infty} (-A_p^{-1} \sum_{m=p}^{\infty} \lambda^{m-p}  A_m)^n A_p^{-1} \vec{b}(\lambda), $$  which series converges to holomorphic function $\vec{\nu}(\lambda) = -  \sum_{n=0}^{\infty} (-A_p^{-1} \sum_{m=p}^{\infty} \lambda^{m-p}  A_m)^n A_p^{-1} \vec{b}(\lambda)$ at the rate of a geometric progression with the denominator $q<1$   for  $|\lambda| \leq \rho.$

Therefore, the load $\vec{x}_{\gamma}(\lambda) = \lambda^{-p} \vec{\nu}(\lambda)$ 
is the Laurent series with $p$th order pole.

Then the following theorem is true.\\

\noindent {\it Theorem 3}.\\
Let $A_0=E,$ $A(\lambda) = \sum_{m=p}^{\infty} A_m \lambda^m,\, p\geq 1.$
Let matrix $A_p$ is not singular. Then there exists  punctured neighborhood 
$0<|\lambda|\leq p$ such as equation  (\ref{eq1}) has the solution $x(t,\lambda)$ 
with pole in point $\lambda = 0$ of order less or equal to $p.$  \\

\noindent {\it Example}.\\
Let us consider the equation
$$ x(t,\lambda) - a(t)x(0,\lambda)  = \lambda \int_0^1 b(t) m(s) x(s,\lambda)\, ds +f(t), \,\, t \in [0,1]. $$

Let us have irregular case of $a(0)=1.$ Let $b(0)\neq 0,$ i.e. condition {\bf I } is not failed,
$$ \frac{d}{d\lambda} A(\lambda)\bigr|_{\lambda=0} = b(0) \int_0^1 m(s) a(s)\, ds. $$
Let
$$ \int_0^1 m(s) a(s) \, ds \neq 0.$$
Then all the conditions of Theorem 3 are fuilfilled for $p=1.$
Then equation has solution $x(\lambda)$ for $|\lambda|>0$
with 1st order pole in point $\lambda=0.$
The desired solution is following
$$ x(t,\lambda) = f(t,\lambda) - \frac{b(t)}{b(0)} f(0) + a(t) x(0,\lambda), $$
where load $x(0,\lambda)$ is constructed as follows
$$ x(0,\lambda) \equiv \frac{1}{(a,m)} \biggl [  - \frac{f(0)}{\lambda b(0)} - (f,m) + \frac{f(0)}{b(0)} (b,m)  \biggr ], $$
where $(a,m) = \int_0^1 a(t) m(t) dt,$
$(f,m) = \int_0^1 f(t) m(t) dt,$
$(b,m) = \int_0^1 b(t) m(t) dt.$
In this example we constructed the solution in the explicit form.


\section{Conclusion and generalizations}

The  linear Fredholm integral functional equations of the second kind with linear functionals is studied. Necessary and sufficient conditions are formulated.
Constructive methods are proposed  for both regular and irregular cases. Typically, the solution is a Taylor series constructed in terms of powers of the parameters. In the irregular case, the solution is constructed as a Laurent series of powers of the parameters. Constructive theory and methods are demonstrated using model examples. 
The case of $A_0 \neq E$ remained not addressed in this paper. The most complete results can be derived for the case of symmetric matrix $A_0$,
In that case solution of equation  (\ref{eq1}) can be also presented as Laurent series 
with pole in point $\lambda=0.$
The corresponding sufficient condition can be derived based on generalized Jordan chains of the theory of perturbed nonlinear operators \cite{Vain}. 
The bifurcation theory of nonlinear loaded integral equations, using the approach of this article in combination with representation theory and group symmetry \cite{log}, 
will also be addressed in future works. Some results in this direction are published in  \cite{Sid23, Sid, Sid1}
The numerical solution of  Fredholm integral-functional equations of the second kind with linear functionals and  parameter will be also addressed in future works.






\end{document}